\documentclass[aps,prl,preprint,superscriptaddress]{revtex4}
\usepackage{graphicx}
\usepackage{color}
\usepackage{portland}
\usepackage{amsmath}
\usepackage{amsfonts}

\begin{document}

\title{Coupling of a high-energy excitation to superconducting quasiparticles in a cuprate from Coherent Charge Fluctuation Spectroscopy.}

\author{B. Mansart}
\affiliation{Laboratory for Ultrafast Microscopy and Electron Scattering, ICMP, Ecole Polytechnique F\'{e}d\'{e}rale de Lausanne, CH-1015 Lausanne, Switzerland}
\affiliation{Laboratory of Ultrafast Spectroscopy, ISIC, Ecole Polytechnique F\'{e}d\'{e}rale de Lausanne, CH-1015 Lausanne, Switzerland}
\author{J. Lorenzana}
\affiliation{Institute for Complex Systems-CNR and Physics Department,
  Sapienza, University of Rome,  Piazzale Aldo Moro 5, I-00185 Rome, Italy}
  \author{A. Mann}
\affiliation{Laboratory for Ultrafast Microscopy and Electron Scattering, ICMP, Ecole Polytechnique F\'{e}d\'{e}rale de Lausanne, CH-1015 Lausanne, Switzerland}  
\author{A. Odeh}
\affiliation{Laboratory of Ultrafast Spectroscopy, ISIC, Ecole Polytechnique F\'{e}d\'{e}rale de Lausanne, CH-1015 Lausanne, Switzerland}  
\author{M. Scarongella}
\affiliation{Laboratory of Ultrafast Spectroscopy, ISIC, Ecole Polytechnique F\'{e}d\'{e}rale de Lausanne, CH-1015 Lausanne, Switzerland}
\author{M. Chergui}
\affiliation{Laboratory of Ultrafast Spectroscopy, ISIC, Ecole Polytechnique F\'{e}d\'{e}rale de Lausanne, CH-1015 Lausanne, Switzerland}
\author{F. Carbone}
\affiliation{Laboratory for Ultrafast Microscopy and Electron Scattering, ICMP, Ecole Polytechnique F\'{e}d\'{e}rale de Lausanne, CH-1015 Lausanne, Switzerland}

\begin{abstract}
Dynamical information on spin degrees of freedom of proteins or solids
can be obtained by Nuclear Magnetic Resonance (NMR) and Electron Spin
Resonance (ESR). A technique with
similar versatility for charge degrees of freedom and their ultrafast
correlations could move forward the understanding of systems like
unconventional superconductors. By perturbing the superconducting
state in a high-$T_c$ cuprate using a femtosecond laser pulse, we
generate coherent oscillations of the Cooper pair condensate which can
be described by an NMR/ESR formalism. The oscillations are detected by
transient broad-band reflectivity and found to resonate at the typical
scale of Mott physics (2.6 eV),  suggesting the existence of a non-retarded contribution to the pairing interaction, as in unconventional (non Migdal-Eliashberg) theories.
\end{abstract}

\maketitle

\section{INTRODUCTION}

According to Bardeen-Cooper-Schrieffer (BCS) theory~\cite{BCS}, superconductivity requires that electrons bind in Cooper pairs and condense collectively in a macroscopic quantum state. In conventional superconductors, the observation of a shift in the superconductivity transition temperature upon isotope substitution~\cite{max50,rey50}, was an experimental breakthrough leading to the conclusion that lattice vibrations (phonons) act as a glue among electrons promoting the required pairing.
 Since the discovery of high temperature superconductivity in cuprates in 1986~\cite{ber1986}, the observation of an analogous fingerprint of the glue involved in the pairing mechanism, if any, has been lacking.

A fertile route to obtain information on excitations in solids and their coupling to electrons is pump-probe spectroscopy~\cite{ore12}. Typically, the sample is illuminated by an ultrashort laser pulse lasting a few tens of fs and carrying 1.5 eV photons. This ``pump'' pulse creates an out-of-equilibrium distribution of particle-hole excitations which decays to states within a few hundreds of meV of the chemical potential~\cite{Perfetti2007} in the pulse duration time scale. There, phase space restrictions slow down the dynamics~\cite{how04} and the subsequent evolution can be studied in real time by a probe pulse. The dynamical response of the system can be observed with a temporal resolution comparable to the time scale of relevant processes in the material, like the pairs breaking, their recombination, or the electron-phonon coupling time.

For example, the photoinduced quenching of the superconducting order parameter and its subsequent recovery were followed by recording the temporal evolution of the gap amplitude in the optical spectrum of different cuprates~\cite{Kaindl2000,sto11,Pashkin2010,Beck2011}. Remarkably, it was found that the energy needed to suppress the superconducting state in these materials is several times larger than the condensation energy~\cite{sto11,Pashkin2010}, in contrast to what happens in conventional superconductors where it is of the same order~\cite{sto11,Beck2011}. Optical studies also provided insights on the relaxation dynamics of the excited  quasiparticles~\cite{Beck2011,ged04,nuh,kab05}, and on the optical spectral weight transfers associated with the carriers kinetic energy changes across the photoinduced phase transition~\cite{Giannetti2011}.

Fs-Angle Resolved PhotoElectron Spectroscopy (ARPES) showed that the decay of photoexcited carriers is dominated by the recombination of the Cooper pairs at the antinodes (i.e. the copper-oxygen bond direction in real space)~\cite{cor11}. Also, similar experiments yielded an estimate of the electron phonon coupling strength being in the intermediate regime~\cite{Perfetti2007}, similar to what was found by fs-electron diffraction, which in turns identified also its anisotropy~\cite{Carbone2008, carbonereview}.

In all of the above-mentioned experiments, excitation by the pump occurs through dipole allowed processes (i.e. the corresponding matter-radiation interaction Hamiltonian is linear as a function of the electric field) and the dynamics is dominated by the incoherent relaxation of the system. However, the same pump pulse can also generate coherent oscillations of the optical properties due to the population of elementary excitations through a stimulated Raman process (i.e. through the second-order term of the matter-radiation Hamiltonian, which is quadratic as a function of the electric field). For example, in transparent media, phonons can be excited by the Impulsive Stimulated Raman Scattering (ISRS) mechanism~\cite{Merlin,Merlin2002}; in absorbing media instead, both a displacive~\cite{Merlin2002,Zeiger,Mazin1994,rif07} and an ISRS mechanism may enter into play depending on various factors such as the lifetime of the particle-hole excitations~\cite{rif07}, the nature of the coupling among phonons and particle-hole excitations, etc.

		\begin{figure}[ht]
		\vspace*{.05in}
\centerline{\includegraphics[width=120mm]{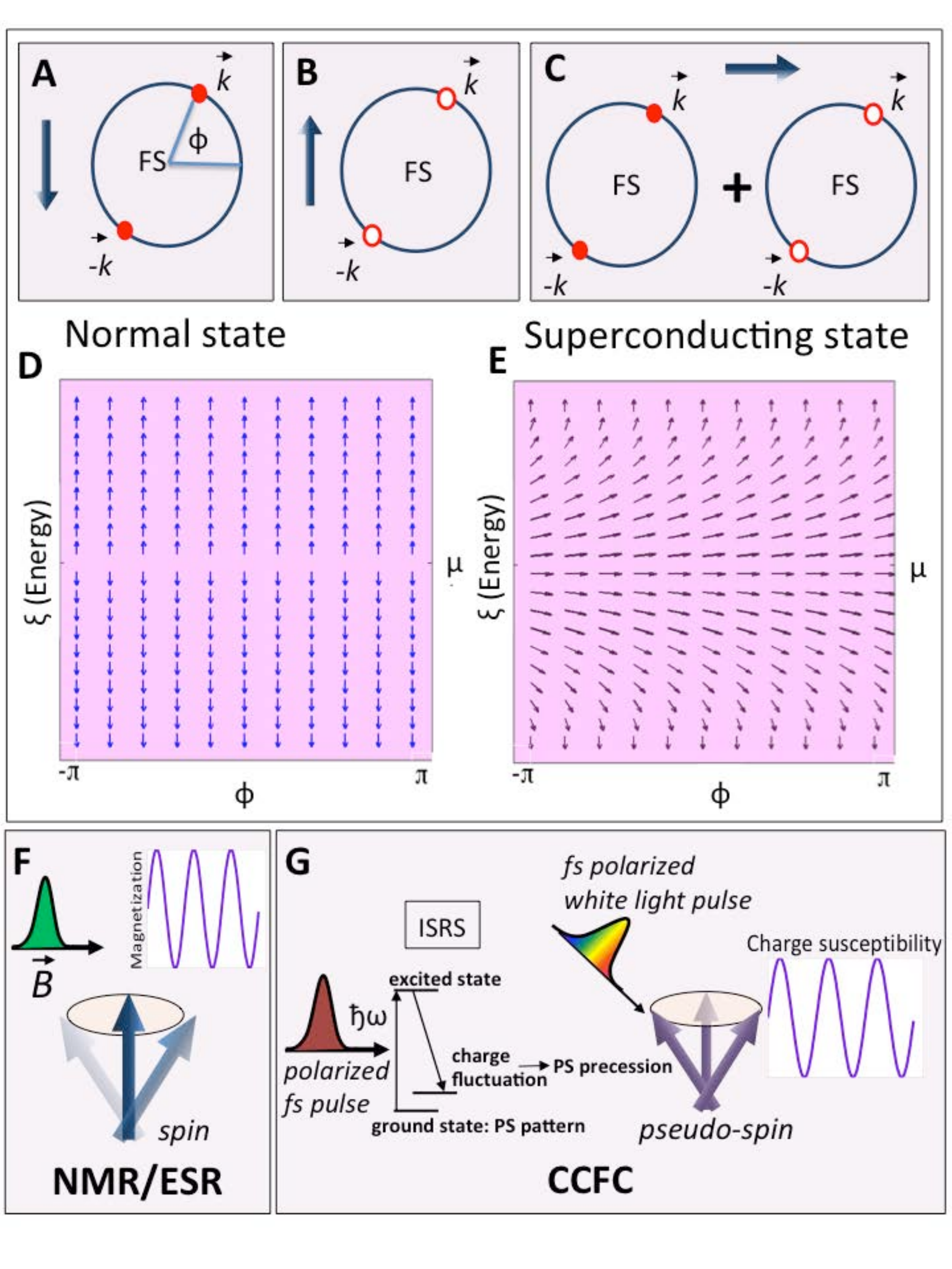}}
\caption{  
Pseudopsin description of the Coherent Charge Fluctuation Spectroscopy experiment. Panel ({\textit A}) defines the angle $\phi$ along the Fermi surface (FS). The three upper panels define the pseudospins operators in momentum space: a pseudopsin down corresponds to the pair of states ($\textbf{k}\uparrow$, $\textbf{-k}\downarrow$) being fully occupied ({\textit A}), 
a pseudospin up to the pair ($\textbf{k}\uparrow$,
$\textbf{-k}\downarrow$) being empty ({\textit B}), and a sideway
pseudopsin to a quantum superposition of the previous two  ({\textit
  C}). Panel ({\textit D}) and ({\textit E}) show the pseudospin
pattern in the normal state and in the case of an $s$-wave
superconductor respectively. Rather than plotting the pseudospins as a
function of momentum $\textbf{k}$ we make a change of coordinates and
plot as a function of the Fermi surface angle  $\phi$ and the energy
distance $\xi$  of the state $\textbf{k}$ from the
chemical potential $\mu$. ({\textit F}) Schematic view of an NMR/ESR experiment in which the spins precess, inducing a magnetization
oscillation, and ({\textit G}) corresponding view for a CCFS experiment, in which the pseudospins precess upon ultrafast excitation and coherent charge fluctuation generation. 
}
\label{pseudo}
\end{figure}

		\begin{figure}[ht]
		\vspace*{.05in}
\centerline{\includegraphics[width=170mm]{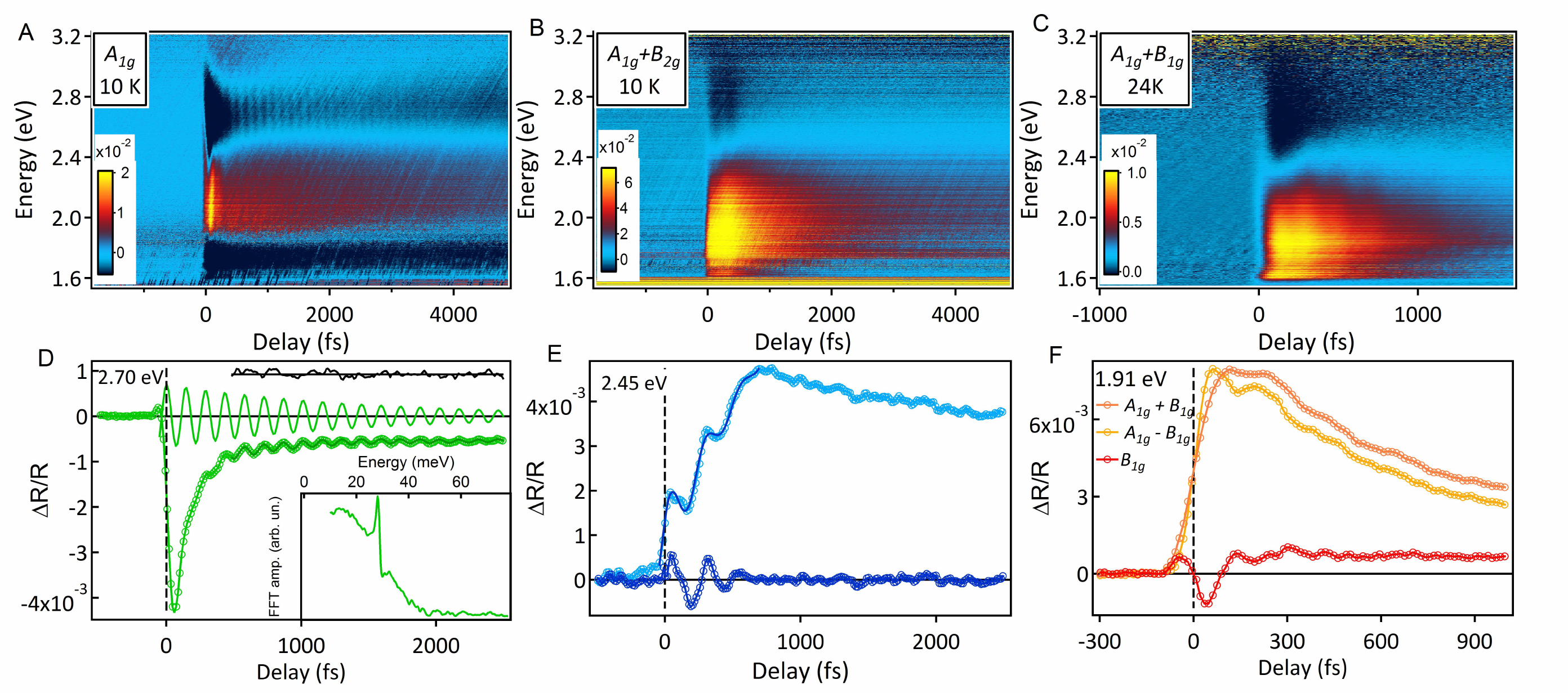}}
\caption{
Transient broad-band reflectivity data at 10 K in $A_{1g}$ [pump $\parallel$ [100], probe  $\parallel$ [001], (\textit{A, D})], $A_{1g}+B_{2g}$ [pump $\parallel$ [110], probe $\parallel$ [110], (\textit{B, E})] and at 24 K $A_{1g}+B_{1g}$ [pump $\parallel$ [100], probe $\parallel$ [100], (\textit{C, F})] geometries (specified in tetragonal axis). The extracted profiles are shown in panels (\textit{C, F}) for selected probe energies. Panel (\textit{E}) presents the reflectivity oscillations by substracting the background on the profile, and in panel (\textit{F}) we show the difference between $A_{1g}+B_{1g}$ and $A_{1g}-B_{1g}$ profiles, which is proportional to the $B_{1g}$ signal. The absorbed pump fluence is around 300 $\mu J/cm^2$.
}
\label{images}
\end{figure}

\begin{figure}[ht]
\vspace*{.05in}
\centerline{\includegraphics[width=120mm]{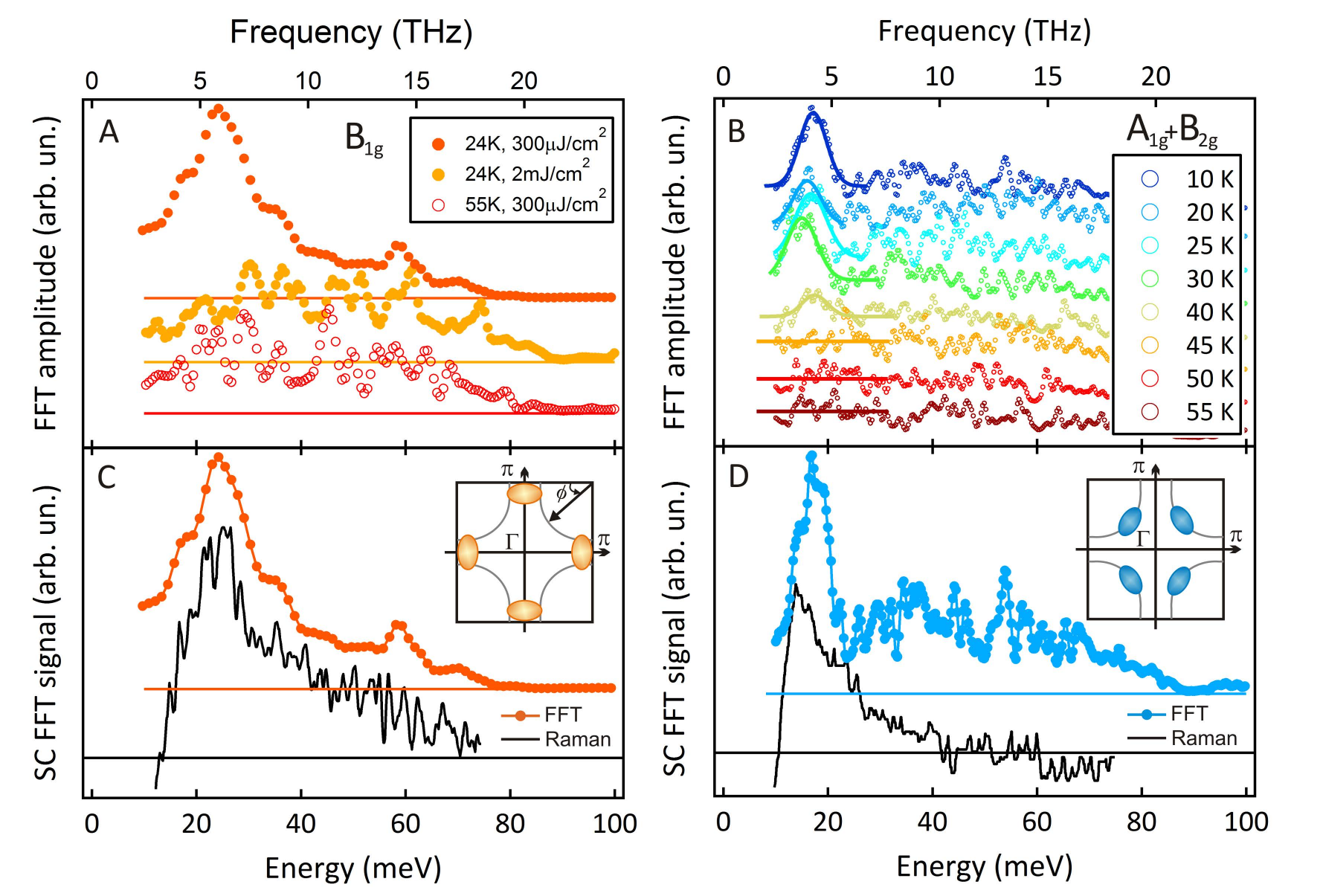}}
\caption{Fourier transform spectra obtained at different temperatures and excitation fluences in $B_{1g}$ ({\textit A}, 1.91 eV probing energy) and $A_{1g}+B_{2g}$ ({\textit B}, 2.45 eV probing energy) geometries; Panels ({\textit C}) and ({\textit D}) show the comparison between transient reflectivity data and Raman measurements, in the superconducting phase. The spontaneous Raman spectra are the difference between superconducting and normal phases, showing only the charge fluctuation peaks. The insets show schematically:  in panel ({\textit C}), the angle $\phi$ along the Fermi surface and shows the regions in momentum space excited in $B_{1g}$ symmetry; panel ({\textit D}), idem for $B_{2g}$ symmetry.}
\label{Raman}
\end{figure}

\begin{figure}[ht]
\vspace*{.05in}
\centerline{\includegraphics[width=125mm]{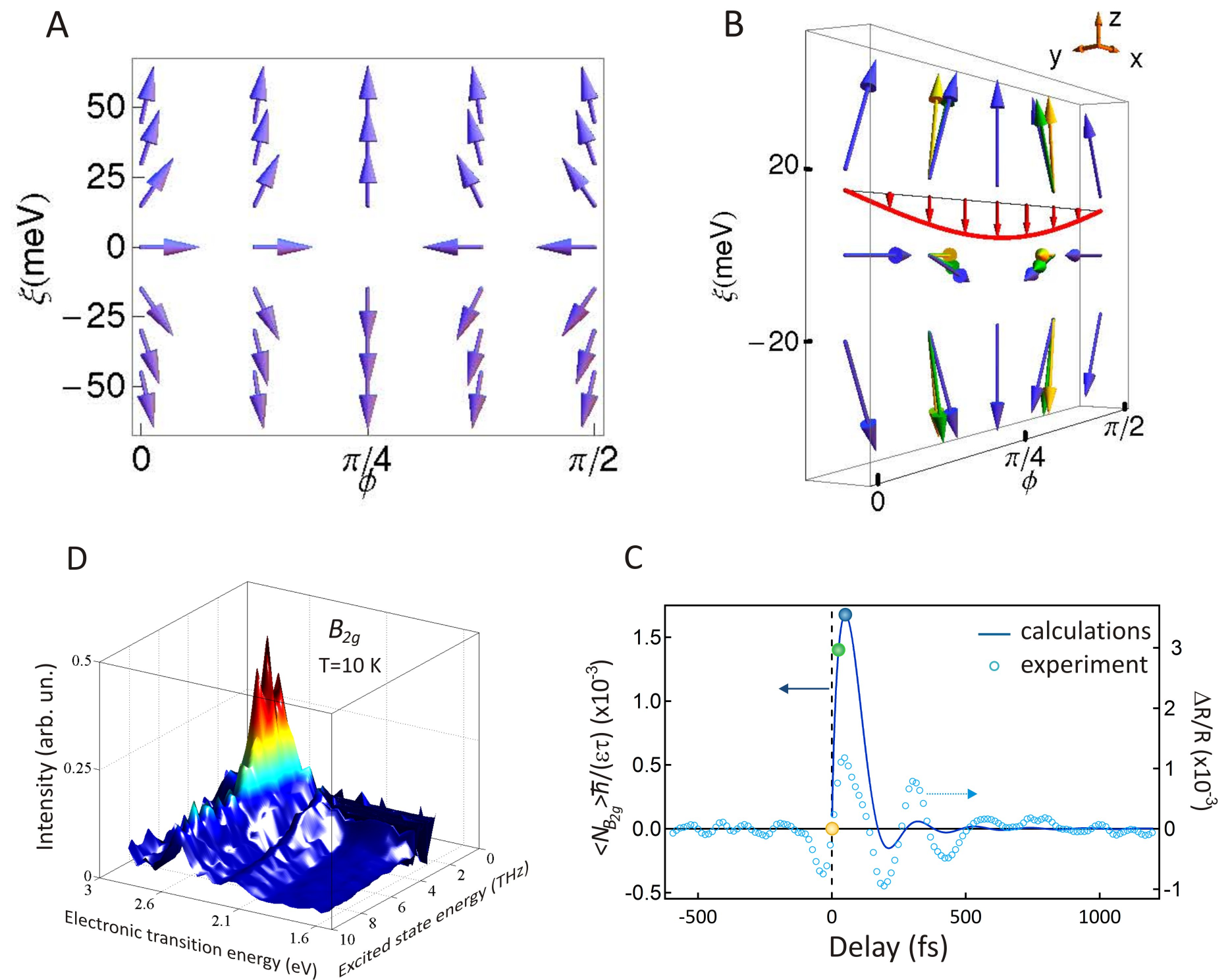}}
\caption{Pseudospin textures coding the BCS wave function in momentum space. ({\textit A}) Ground state texture: pseudospins are labeled by the distance in energy from the Fermi level $\xi\equiv\xi_{\bf k}$ and the angle $\phi$ along the Fermi surface (inset of Fig. 2 C); ({\textit B}) Small red arrows: amplitude of the impulsive field $\delta{\bf b}_{\bf k}$ applied at $t=0$ in $B_{2g}$ symmetry. Long arrows: texture snapshots (amplitudes exaggerated for clarity) immediately after the excitation (yellow), at 25~fs (green) and at 50~fs (blue); ({\textit C}) Theoretical charge fluctuation: solid dots corresponds to the snapshots of panel ({\textit B}). The open dots are the experimental change in reflectivity after the high energy $p-h$ background has been subtracted; ({\textit D}) Probe energy dependence of the Fourier transformed $A_{1g}+B_{2g}$ fluctuation.}
\label{texture}
\end{figure}

\begin{figure}[ht]
\vspace*{.05in}
\centerline{\includegraphics[width=100mm]{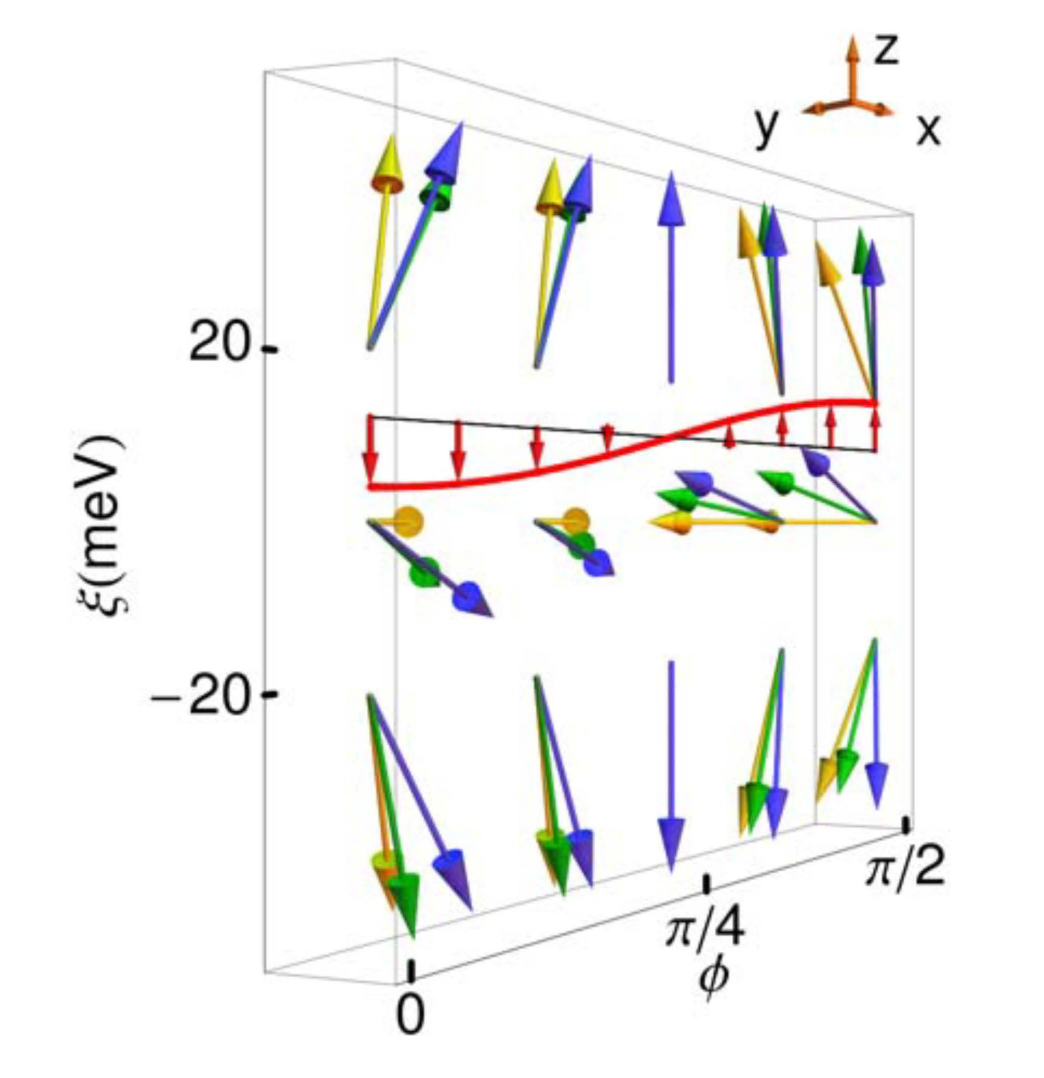}}
\caption{Snapshots of the textures coding the BCS wave function in momentum space. Small red arrows: amplitude of the impulsive field $\delta{\bf b}_{\bf k}$ applied at $t=0$ in $B_{1g}$ symmetry. Long arrows: texture snapshots (amplitudes exaggerated for clarity) immediately after the excitation (yellow), at $25$ fs (green) and at $50$ fs (blue).}
\label{fig:b1gtext}
\end{figure}

\begin{figure}[ht]
\vspace*{.05in}
\centerline{\includegraphics[width=100mm]{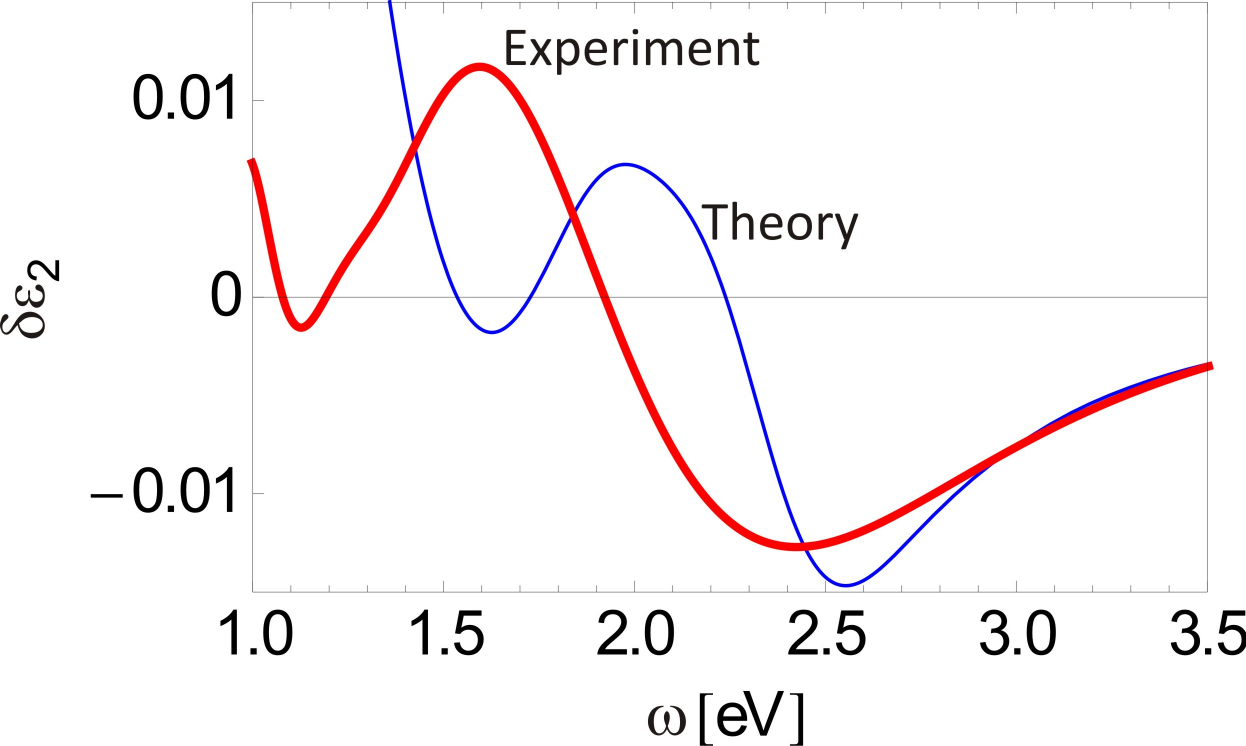}}
\caption{Comparison between the experimental (red thick line) and theoretical (blue thin line) transient imaginary part of the in-plane component of the dielectric function tensor
 $\varepsilon_2$ at 500 fs time delay for doping $x=0.15$. For the theory we used  $\delta n_{CT}(500 fs)=8\cdot10^{-3}$ and three band parameters as specified in the text.}
\label{fig:deps2dnct}
\end{figure}

In cuprates, the generation of coherent structural excitations (phonons) has already been reported~\cite{Albrecht1992,mis02}. More in general, any excitation that is active in spontaneous Raman scattering is also allowed in pump-probe spectroscopies. Indeed, Raman active electronic excitations like magnons and density fluctuations have been found to be coherently generated in different materials~\cite{zha04, bao04}.  

 In this report, we present a technique, Coherent
  Charge Fluctuation Spectroscopy (CCFS), in which charge fluctuations
  are coherently generated by the pump pulse through a stimulated
  Raman process active in a superconductor. These coherent excitations are subsequently probed by
  fs-broad-band reflectivity, allowing to observe the real-time
  oscillations of the Cooper pair condensate and its impact on high-energy excitations. CCFS takes advantage of
the possibility to coherently control Cooper pairs in superconductors
in a way that resembles the coherent control of spins with NMR-ESR
techniques. 
 To understand this analogy, it is useful to use Anderson's pseudospins formalism~\cite{and58,war05,pseudospins_ref}.
The latter is based on the fact that despite their obvious physical difference, from a mathematical (or purely formal) point of view, magnetism and superconductivity are closely linked phenomena.

The BCS wave function of a generic superconductor reads: 
\begin{equation}
 \label{eq:hbcs1}
\left| \Psi\right\rangle=\prod_{k} (u_{\textbf{k}} + v_{\textbf{k}} c^{\dagger}_{\textbf{k} \uparrow} c^{\dagger}_{\textbf{-k} \downarrow}) \left| 0 \right\rangle
\end{equation}
where the operator $c^{\dagger}_{\textbf{k} \sigma}$ creates an electron with spin $\sigma$ and wavevector $\textbf{k}$ and $\left| 0 \right\rangle$ represents the vacuum state. For each pair of states ($\textbf{k}\uparrow$, $\textbf{-k}\downarrow$) the wave function is a quantum mixture of the pair being empty (with amplitude $u_{\textbf{k}}$) and being fully occupied (with amplitude $v_{\textbf{k}}$). Anderson's idea is illustrated pictorially in Fig.~\ref{pseudo} and consists of representing the fully occupied pair ($\textbf{k}\uparrow$, $\textbf{-k}\downarrow$) by a down pseudospin in momentum space (\textit{A}) and the state in which the pair is empty by an up pseudospin (\textit{B}) (see also Ref.~\cite{pseudospins_ref}). The advantage of this representation is that the pseudospins behave like traditional spin-1/2 operators, and the quantum mixture of fully occupied and empty states in the BCS wave function is represented by a sidewise pseudospin (Fig.~\ref{pseudo}~\textit{C}).

 In the normal state $u_{\textbf{k}}=0$ and
  $v_{\textbf{k}}$ is different from zero only for states inside the
  Fermi surface, which corresponds to empty pairs above the chemical
  potential (up pseudospins) and fully occupied pairs below the
  chemical potential (down pseudospins), leading to the pseudospin texture shown schematically in Fig.~\ref{pseudo}~\textit{D}, with a sharp interface at the Fermi surface. In the superconducting state, mixing of empty and fully occupied pairs, which becomes maximum at the chemical potential, blurs the Fermi surface, leading to the texture shown in panel \textit{E}.

Coherent control of the pseudospins in a superconductor can be
achieved by an {\em ad hoc} prepared light pulse through a stimulated
Raman process which, as discuss in more detail below, triggers the
precession of the pseudospins around their equilibrium axis. This is
analogous to NMR and ESR experiments in which magnetic field pulses
induce a precession of real spins~\cite{sch96}. The concept and
schematics of this experiment are depicted in Fig.~\ref{pseudo}
\textit{F} and \textit{G}. An infrared polarized fs laser pulse
couples to charge fluctuations in a superconductor according to Raman
selection rules. The pump pulse impulsively perturbs the system and
induces the pseudospins precession, i.e. the oscillations of the
Cooper pair condensate. The optical spectra of the system are then
monitored in real time at different energies, revealing the optical
transitions that respond to the oscillating condensate; this allows to
single out those excitations that can potentially mediate electron-electron interactions
impacting the formation of Cooper pairs.
This is of pivotal importance for cuprates, since the applicability of
conventional pairing theories~\cite{eli60}, based on retarded
interactions between electrons mediated by low energy glue bosons, has
been doubted~\cite{and07,zaa11} and a completely different framework
has been proposed involving non-retarded interactions associated with 
electronic high-energy scales~\cite{and87}.

\section{EXPERIMENTS}

We performed such high-temporal resolution ($< 50$~fs) experiments in two optimally doped ($T_c$ = 40 K) La$_{2-x}$Sr$_x$CuO$_4$ (LSCO) single crystals ($x=0.15$) with different orientations (see \textit{SI Text} and Ref.~\cite{lu} for details). A polarized 1.55 eV laser pulse with a duration of 45 fs and an absorbed fluence around 300 $\mu J/cm^2$ (unless otherwise stated) induces both dipole (linear in the electric field) and Raman (quadratic in the electric field) allowed excitations, the latter being the main focus of this work. 

We chose different experimental geometries for exploiting the Raman selection rules for excitation and detection to obtain information on different final states (see \textit{SI Text}). In the first geometry, the pump electric field is parallel to the Cu-O bond giving access to Raman excitations with $A_{1g}+B_{1g}$ symmetry, while the probe pulse electric field is directed towards the $c$-axis, allowing us to detect only $A_{1g}$ symmetry excitations. Then, using the same pumping geometry, we probed the excited system along $\left[100\right]$ and $\left[010\right]$, which respectively give access to $A_{1g}+B_{1g}$ and $A_{1g}-B_{1g}$ excitations. Performing the difference between the two orientations allows us to extract only $B_{1g}$ excitations. Finally, we used the pumping and probing fields on the diagonal direction giving access to $A_{1g}+B_{2g}$ Raman excitations. The dynamics of all these excitations is then probed by broad-band ultrafast reflectivity, which overall time-energy dependence is displayed in Fig.~\ref{images}~\textit{A-C}.

The transient reflectivity is dominated by a large abrupt amplitude change followed by a relaxation; this is a consequence of high energy particle-hole ($p$-$h$) excitations produced by the dipole allowed absorption of the pump photons. Furthermore, in both orientations the transient reflectivity changes sign throughout the spectra in correspondence to specific electronic transitions. These changes reflect the transfer of spectral weight among the different absorption bands produced by the  $p$-$h$ excitations~\cite{Giannetti2011}. The number of $p$-$h$ excitations involved is estimated in the \textit{SI Text} to be less than $ 10^{-2}$ per Cu atom (Fig.~S2).

All geometries present coherent oscillations of Raman excitations on top of the dipole $p$-$h$ excitations relaxation as shown in the temporal profiles in Fig.~\ref{images}~\textit{D-F}, taken at selected energies where the oscillation amplitude is the largest. These profiles are representative of the full data sets for a given geometry as far as the oscillation phase and frequency are concerned. In $A_{1g}$ symmetry, an ultrafast oscillation with a period of 145~fs and a long coherence time (1.45 ps) is visible at all wavelengths (Fig.~\ref{images}~\textit{A} and \textit{D}).
 The Fourier transform analysis of $A_{1g}$ symmetry data is presented as an inset in Fig.~\ref{images}~\textit{D}. A sharp peak at 28~meV is visible, corresponding to the out-of-plane La $A_{1g}$ mode of LSCO~\cite{Sugai2003}; the corresponding atomic motions are shown in Movie S1. Such coherent fully symmetric modes have already been observed in high-$T_c$ superconductors~\cite{Mazin1994, Albrecht1992}.
 
Instead, in both $A_{1g}+B_{2g}$ and $B_{1g}$ symmetries (Fig.~\ref{images}~\textit{E} and difference in Fig.~\ref{images}~\textit{F}), slower and damped (around 300~fs coherence time) oscillations are clearly observed below $T_c$. The Fourier analysis of these time-resolved profiles is presented in Fig.~\ref{Raman}~\textit{A-B}. In $A_{1g}+B_{2g}$ symmetry, for a probing wavelength of 2.45 eV, the temperature dependence of the Fourier-transform signal shows an obvious peak at 18~meV that vanishes above $T_c$ (Fig.~\ref{Raman}~\textit{B}). In $B_{1g}$ symmetry, the broad peak appears at energies around 24~meV when the sample temperature is lower than $T_c$. Increasing the pump fluence to 2~$mJ/cm^2$, no such peak could be observed below $T_c$ (Fig.~\ref{Raman}~\textit{A}).

In Fig.~\ref{Raman} \textit{C-D} we display the $THz$ spectra obtained in the superconducting state, and compare them with the spontaneous Raman response (data taken from~\cite{Sugai2003}), which is well understood in terms of the excitation of two Bogoliubov quasiparticles~\cite{dev07}. The good agreement between them allows us to identify the strongly temperature dependent part of the oscillations as Raman charge fluctuations of the superconducting condensate. This agreement is expected from simple theoretical considerations for electronic ISRS which show that any excitation that is Raman active in a colinear configuration of incoming and outgoing photon electric field is also accessible in a pump-probe experiment (see \textit{SI Text}). Our experiment detects remnants of superconductivity at fluences of the same order but larger than previously reported~\cite{cor11, Pashkin2010}. We attribute that difference to the much higher sensitivity of our optical measurement to superconductivity and its bulk character. Presumably, superconductivity is indeed quenched on the first layers of the sample and becomes invisible to surface probes like photoelectron spectroscopies~\cite{cor11}.

The temporal evolution of the coherent phonon oscillation in $A_{1g}$ geometry is presented in Fig.~\ref{images}~\textit{D}, with its extrapolation down to zero-time delay (as defined in~\cite{lu}), allowing us to establish its cosine waveform which is typical of a displacive (resonant) mechanism of excitation~\cite{Merlin2002, Zeiger}. The electronic transitions induced by 1.55 eV photons occurs between the ground state of the material and higher energy electronic states. At this energy, a peak in the optical absorption is observed in LSCO, distinct from the Cu-O charge transfer~\cite{Uchida1991}, which had been attributed to charge ordering in the form of stripes~\cite{Lorenzana2003}. Thus, the cosine wave form indicates that, not unexpectedly, the charge modulations are strongly coupled to the $A_{1g}$ La phonon.

In the $A_{1g}$ geometry, the presence of the strong coherent phonon disturbs the real-time observation of the superconducting condensate. Instead, in both $A_{1g}+B_{2g}$ and $B_{1g}$ symmetries, the fluctuations of the superconducting quasiparticles are clearly
observable and start at zero time delay, allowing the determination of a sine waveform (Fig.~\ref{images} \textit{E-F}). This indicates that contrary to the $A_{1g}$ phonon case, the triggering mechanism is ISRS~\cite{Merlin,Merlin2002}, meaning that the $p$-$h$ excitations at the energy of the pump pulse are not directly coupled to the superconducting quasiparticles. We show below that an analysis of the probe energy dependence leads to the same conclusion.

\section{DISCUSSION}
For the Raman allowed excitations, the effect of the pump light on the electrons can be described by a time-dependent impulsive potential
quadratic in the electric field (see \textit{SI Text} and Ref.~\cite{raman}). As mentioned above, we can describe its effect on the superconducting quasiparticles using Anderson's pseudospin formalism \cite{and58}. 

The reduced BCS Hamiltonian in the presence of a time-dependent potential acquires a simple form when written in term of the pseudospin operators ${\boldsymbol \sigma}_{\bf k}$~\cite{pseudospins_ref},
\begin{equation}
 \label{eq:hbcs}
H_{BCS}=-\sum_{\bf k}{\bf b}_{\bf k}.{\boldsymbol \sigma}_{\bf k},
\end{equation}
here ${\boldsymbol \sigma}_{\bf k}$ is a Pauli matrix representing the 
pseudospin associated with the pair of states ($\textbf{k}\uparrow$,
$\textbf{-k}\downarrow$) and ${\bf b}_{\bf k}$ is a fictitious
``magnetic field''. At the equilibrium, pseudospins orient parallel to 
the ground state pseudomagnetic field ${\bf b}^0_{\bf k}=(\Delta_{\bf k},0,\xi_{\bf k})$, where $\Delta_{\bf k}$ is the superconducting order parameter and $\xi_{\bf k}=\epsilon_{\bf k}-\mu$, $\epsilon_{\bf k}$ being the quasiparticle band energy (in the absence of superconductivity)  and $\mu$ is  the chemical potential. Thus, this Hamiltonian express the familiar fact that the ground state wave function is determined by the mean-field order parameter $\Delta_{\bf k}$, which in turn can be expressed in terms of the pseudospins. 

In the absence of superconductivity, $\Delta_{\bf k}=0$; so the pseudomagnetic field points in the $z$ direction and 
changes sign at the chemical potential, leading to the equilibrium texture of  Fig.~\ref{pseudo} \textit{D}. In the superconducting state, the pseudomagnetic field acquires a horizontal component, $\Delta_{\bf k} \neq 0$, so that in the case of an $s$-wave superconductor the pseudospins display the texture shown in  Fig.~\ref{pseudo} \textit{E}. For a $d$-wave superconductor, the horizontal component of the pseudomagnetic field cancels along the nodal directions due to the gap anisotropy, leading to the texture of Fig.~\ref{texture} \textit{A} which has no sideway pseudospins along the nodal direction.

 The Raman coupling to the pump pulse can be
  described by a time-dependent potential $v^X_{\bf k}(t)$ coupling
  to charge fluctuations, i.e. to the z-component of the pseudospins. The
  potential has a different dependence in momentum space depending on
  the symmetry $X=A_{1g}$, $B_{1g}$, $B_{2g}$ which is determined by
  the polarization of the pump  (see \textit{SI Text} and Ref.~\cite{pseudospins_ref,raman}).  Thus the pseudomagnetic field becomes time dependent: ${\bf b}_{\bf k}(t)={\bf b}^0_{\bf k}+\delta{\bf b}_{\bf k}(t)$, with $\delta{\bf b}_{\bf k}(t)=(0,0,v^X_{\bf k}(t))$. The pseudospins obey the usual equations of motion for magnetic moments in a time-dependent magnetic field~\cite{and58},
\begin{equation}
 \label{eq:motion}
\hbar \frac{\partial\boldsymbol {\sigma}_{\bf k}}{\partial t} =-2 [{\bf b}^0_{\bf k}+\delta{\bf b}_{\bf k}(t)]\times {\boldsymbol \sigma}_{\bf k}.
\end{equation}
implying that after the pulse passage the pseudospins precess around the equilibrium direction with an angular velocity $2|{\bf b}^0_{\bf k}|/\hbar$,
$|{\bf b}^0_{\bf k}|=\sqrt{\xi_{\bf k}^2+\Delta_{\bf k}^2}$ being the BCS quasiparticle energy (see Fig.~\ref{texture}~\textit{B}, Fig. S1
and Movie S2).

Equation~(\ref{eq:motion}) is close to the equation of motion used
in NMR/ESR formalisms; however, in NMR the static field ${\bf b}^0$ is usually
provided by an external field, whereas here it is due to the
interaction with the other pseudospins. The magnetic 
analogy is actually more complete with ESR in magnetically ordered
materials where ${\bf b}^0$ can be completely due to the interaction with the other spins.

Since $\delta{\bf b}_{\bf k}$ is in the $z$ direction, only
pseudospins having a significant component at equilibrium in the
$x$-$y$ plane respond to the Raman impulsive field, automatically
selecting the quasiparticles participating in the pairing. This is
further constrained by the momentum dependent form factors in
$v^X_{\bf k}(t)$~\cite{raman}.
Therefore, in $B_{2g}$ symmetry only pseudospins which are close to the Fermi level and are neither in the nodes nor in the antinodes have a significant
time dependence. The oscillating $z$ projection of the pseudospins shown in Fig.~\ref{texture}~\textit{B} encodes the contribution to the total
charge fluctuation, shown in panel \textit{C}. At $t=50$~fs (blue) the pseudospins at $\xi=0$ and close to $\phi=\pi/4\pm \pi/8$ are close to their maximum negative amplitude in the $z$ direction, corresponding to the first peak in the $B_{2g}$ charge fluctuation.

We also show in Fig.~\ref{texture} \textit{C} a comparison between the experimental and theoretical condensate oscillation in $B_{2g}$ geometry. Interestingly, the experiment shows a quite long coherence time compared to theory. Details of the computations are given in the \textit{SI Text}.

The transient optical properties of the system in the presence of a fluctuation of symmetry $X$ are governed by the changes in the dielectric function tensor 
\begin{equation}
  \label{eq:tdc}
  \delta
\boldsymbol{\epsilon}(\omega,t)=- 4\pi \sum_{X}\frac{\partial{\boldsymbol{\chi}}}{\partial \langle N_X \rangle }(\omega) \langle N_X \rangle(t),
\end{equation}
and $\partial{\boldsymbol{\chi}}/{\partial
 \langle N_X \rangle}$ is the conventional Raman tensor (see \textit{SI Text}). We see that the same Raman tensor appears in the generation of the pulse by ISRS and in the subsequent probing process. In analogy with lattice ISRS~\cite{Merlin2002}, only excitations having an interaction matrix element with the fluctuating quasiparticles will contribute to ${\partial{\boldsymbol{\chi}}}/{ \langle N_X \rangle}$ allowing to detect excitations participating in the pairing. We point out that CCFS is not restricted to reflectivity, and other techniques like spontaneous Raman scattering can be used as a probe allowing to test also excitations of different symmetries. In this case a different matrix element will be involved in the probe in lieu of the Raman tensor in Eq.~(\ref{eq:tdc}).

 The oscillation of the superconducting condensate is most clearly visible in the $A_{1g}+B_{2g}$ configuration. For this reason, we perform the spectral analysis in this configuration. The probe-energy dependence of the $A_{1g}+B_{2g}$ fluctuation in the frequency domain is presented in Fig.~\ref{texture} \textit{D}. The superconducting fluctuations clearly resonate at an energy of 2.6 eV,
corresponding to the Cu-O charge transfer energy of the parent
compound which coincides with the Hubbard energy $U$ of a one-band
description~\cite{and07}. Remarkably, even though there is substantial
absorption below the charge transfer band in our samples, the
superconducting quasiparticles appear to be decoupled from the
excitations in that energy region. This is fully
consistent with our finding above that the $A_{1g}+B_{2g}$
fluctuations have a sine waveform when pumped at 1.55 eV.

The correct framework to understand superconductivity in cuprates has been subject of an intense debate~\cite{and07, zaa11, mai08}. One
possibility is that the role of phonons in the traditional mechanism is replaced by a different low energy bosonic excitation like damped
magnons which act as a glue allowing the pairing of electrons~\cite{sca95,pin97}. In this scenario, superconductivity can be understood in the traditional framework~\cite{eli60} where retardation plays an important role. Anderson~\cite{and07} has argued that there is no such a low energy glue and that proximity to the Mott phase is an essential ingredient. The relevant time scale of the interactions inducing the pairing is the inverse of the Hubbard energy $U\gtrsim 2eV$. Therefore, the interaction can be considered instantaneous for practical purposes. Our results are consistent with a coupling of the superconducting quasiparticles with excitations at 2.6 eV. We attribute this to a fingerprint of ``Mottness'' in the superconducting state, although we can not exclude other electronic transitions like a $d-d$ exciton which would also be interesting. Systematic studies of this type of oscillations in different chemical compositions and energy ranges coupled to further theoretical work may allow to have deeper insights in the pair-mediating or pair-breaking nature of these excitations.

A negligible coupling in the rest of the measured energy window (1.6 eV $<\hbar\omega< $ 3.2 eV) is observed, but we cannot exclude that
other electronic excitations outside our probing range are also coupled to superconductivity and even dominant. Numerical computations
support a coupling to the Mott scale~\cite{mai08}, although with a strong contribution from the low energy region.

The key feature of the isotope effect~\cite{max50,rey50} in conventional superconductors was its high specificity, since only the frequency of one potential glue excitation was affected and its impact on superconductivity evaluated. CCFS has a high degree of specificity in a reverse form: only paired electrons are affected, and their impact on different excitations assessed. 

Compared to previous ultrafast studies of
  superconductivity, our experiments provide a direct observable of
  the coherent Cooper pairs dynamics. Moreover, because of the spectroscopic nature of our probing scheme, we can detect resonances between superconductivity and high-energy excitations. Also, because we directly obtain the condensate oscillations in real time, we have access to their phase and its evolution throughout the probing energy range. The presented results form a benchmark for time-resolved experiments in cuprates and shed new light on the nature of the pairing interactions.

In a more general perspective the NMR/ESR analogy encoded in Eqs.~(\ref{eq:hbcs}) and (\ref{eq:motion}) allows to borrow concepts like the relaxation times $T_1$ and $T_2^*$~\cite{sch96}. $T_2^*$ is defined by the decay of the charge fluctuations, which is dominated by the inhomogeneity of the pseudomagnetic field in momentum space. Therefore, our experiment opens appealing perspectives to typical NMR/ESR-like techniques such as coherent control of the superconducting wave-function by a sequence of pulses. These tools can be generally applied to different materials including heavy fermions and iron-based superconductors.

\acknowledgements
The authors acknowledge useful discussions with A.B. Kuzmenko and D. Fausti. This work was supported by the Swiss NSF via the contracts $PP00P2-128269$ and $20020-127231/1$.
J. Lorenzana is supported by Italian Institute of Technology-Seed project NEWDFESCM.

\appendix

\section{Experiments}

Two single crystals of optimally doped LSCO were cut, polished and oriented via X-ray diffraction to obtain two surfaces containing respectively the $a$ and $b$ or the $a$ and $c$ crystallographic directions.

Pump-probe reflectivity was performed with a 1 kHz Ti:Saph amplified fs laser which output was splitted between a monochromatic (1.55 eV) pump beam, capable of fluences up to 10 mJ/cm$^2$~\cite{lu} and a probe beam. The latter was used to generate a white light continuum (1.6-3.2 eV) pulse by passing through a CaF$_2$ nonlinear crystal. The probe white light pulses were collected and focused on the sample surface by two parabolic mirrors. The samples were placed in a closed-cycle cryostat allowing measurements from room temperature down to 10 K at a pressure of $10^{-8}~mbar$.

 The reflectivity spectrum and a reference signal were synchronously detected by two identical spectrometers composed by dispersing gratings and photodiode array detectors performing single-shot aquisition.

 The pulse duration is 45 fs which puts a lower limit~\cite{Merlin} to the frequency of the excitations that can be excited with Impulsive Stimulated Raman Scattering (ISRS), $\omega> 1/(45 fs)$.

 The experimental setup scheme as well as further informations may be found in Ref.~\cite{lu}.

\section{Theory}

\subsection*{Coherent generation of Cooper-pair condensate charge fluctuations}
\ \

\label{sec:coher-gener-coop}

In this section, we discuss the computation of the Raman charge fluctuation in the superconducting state.
Within a one-band description of electrons close to the Fermi surface we consider uniform ({\it i.e.} zero momentum) charge fluctuations described by the operator~\cite{dev07},

$$N_X=\sum_{{\bf k}\sigma} f_{\bf k}^X n_{{\bf k}\sigma}.
$$
where $X$ runs over the possible Raman symmetries: $f_{\bf k}^{A_{1g}}=[\cos(k_x a)+\cos(k_y a)]/2$, $f_{\bf k}^{B_{1g}}=[\cos(k_x a)-\cos(k_y a)]/2$, $f_{\bf k}^{B_{2g}}=\sin(k_x a)\sin(k_y a)$. $n_{{\bf k}\sigma}=c^\dagger_{{\bf k}\sigma} c_{{\bf k}\sigma}$ is the occupation operator for the state with wavevector ${\bf  k}$ and spin $\sigma$ and  $c^\dagger_{{\bf k}\sigma}$ ($c_{{\bf k}\sigma}$) are creation (destruction) operators for electrons.

Generalizing the arguments of Ref.~\cite{Merlin,Merlin2002} for lattice ISRS to electronic ISRS, we write the Hamiltonian of the system in the presence of the pump pulse as,
$
H=H_{BCS}+H_R
$
where 
$$H_{BCS}=\sum_k \xi_k n_{{\bf k}\sigma}-
\sum_k(\Delta_k^*  c^{\dagger}_{-k\downarrow}c^{\dagger}_{k\uparrow} +h.c.)
$$
is the BCS reduced Hamiltonian describing the low-energy
superconducting quasiparticles,  
with $\xi_k=\epsilon_k-\mu$ the band energy measured from the Fermi
level, $\Delta_k$  the $d$-wave superconducting order parameter and
$$
H_R=\sum_X v_X(t) N_X
$$
is the perturbation due to the pump laser.

In the semiclassical approximation and for $\omega_L$ much larger than the frequency of the fluctuations,
\begin{equation}
  \label{eq:vx}
 v_X(t)=-\frac12{\bf E}(t).\frac{\partial {\boldsymbol
     \chi}(\omega_{L}) }{\partial \langle N_X \rangle}.{\bf E}(t).
\tag{S1}
  \end{equation}
Here ${\boldsymbol \chi}$ is the charge susceptibility,
$\partial {\boldsymbol \chi}(\omega_{L}) /\partial \langle
  N_X \rangle $
is the conventional second rank Raman tensor for electronic scattering with symmetry $X$ and incident frequency $\omega_L$~\cite{Cardona}, ${\bf E}$ is the time dependent electric field of the pump wave and $X$ runs over the allowed symmetries.

In spontaneous Raman scattering the operator has the same form as in
Eq.~(\ref{eq:vx}), unless one can change independently the electric
fields $\textbf{E}(t)$ on the left and on the right of the Raman
tensor . Thus the selection rules for ISRS and spontaneous Raman are
quite similar. Disregarding orthorombicity and  using the symmetry
properties of the Raman tensor in LSCO~\cite{dev07} ($D_{4h}$ group)
for a pump pulse polarized in the $(a,b)$ plane the Raman operator
reads
\begin{equation}
H_R =-\frac{E(t)^2}2 \left[ \frac{\partial \chi_{xx}}{\partial
     \langle N_{A_{1g}}\rangle}[(\hat e_x)^2+(\hat
   e_y)^2]N_{A_{1g}}\right.\nonumber
\end{equation}
\begin{equation}
+\left. \frac{\partial \chi_{xx}}{\partial \langle N_{B_{1g}}\rangle}(\hat e_x \hat e_x-\hat e_y \hat e_y)N_{B_{1g}}  
+ \frac{\partial \chi_{xy}}{\partial \langle N_{B_{2g}}\rangle}2\hat e_x \hat e_y N_{B_{2g}}\right]\tag1{S2} 
\end{equation} 
with $\hat {\bf e}$ a versor in the direction of the electric field (${\bf E}=E \hat {\bf e}$) and the $x$-axis parallel to the CuO bond direction. We see that for a pulse in the [100] direction, $A_{1g}+B_{1g}$ symmetries are excited  [Fig.~2 {\it A} and {\it C}]
while if the electric field is aligned along [110], $A_{1g}+B_{2g}$ symmetries are excited [Fig.~2 {\it B}]. Selection rules for the detection can be derived analogously from Eq.~(3). In the case of Fig.~2 {\it A} the probe field is in the [001] direction and
only the $A_{1g}$ component of the $A_{1g}+B_{1g}$ fluctuation can be seen.

Using linear response for symmetry $X$ the fluctuations at zero
temperature can be obtained as, 
\begin{equation}
  \label{eq:lehman}
\langle  N_X\rangle(t)=- \int_{-\infty}^tdt'\sum_\nu \sin[\omega_\nu
(t-t')]|\langle 0| N_X |\nu \rangle|^2  v_X(t')
\tag{S3}
\end{equation}
with the sum running over a complete set of states, and $\left|0\right\rangle$ is the ground state before excitation and $\left|\nu\right\rangle$ the excited one with presence of charge fluctuations $N_X$. If the pump pulse of width $\tau$ and amplitude $\epsilon$ is approximated  by a Dirac delta function ($v_X(t)=\delta(t) \tau \epsilon$), Eq.~(\ref{eq:lehman}) yields a sine like wave form with shape determined by the Fourier transform of the conventional Raman scattering line shape for symmetry $X$~\cite{dev07}.

It is also instructive to compute the dynamics of the charge fluctuation in a $d$-wave BCS state using pseudospins operators defined as 
$${\sigma}^x_{\bf k}=(c_{{\bf k}\uparrow}c_{-{\bf k}\downarrow} +h.c.),$$
$$i{\sigma}^y_{\bf k}=(c_{{\bf k}\uparrow}c_{-{\bf k}\downarrow} -h.c.),$$
$${\sigma}^z_{\bf k}=1-n_{{\bf k}\uparrow}-n_{-{\bf
    k}\downarrow},$$
and which allow to rewrite $H_{BCS}$ in terms of
  pseudospins in a pseudomagnetic field (Eq. [2]). 
 Also  using the definitions in the main text (Ref. [41]) we rewrite $H_R$ as
 a time dependent contribution to the  pseudomagnetic field 
 in the $z$ direction of magnitude  $v^X_{\bf k}(t)=v_X(t)
  f^X_{\bf k}$.

We first linearize the equation of motion [Eq. (3)] for the pseudospins~\cite{and58, Parmenter} 
in terms of the time dependent fluctuations $\delta {\boldsymbol \sigma}_{\bf k}(t)\equiv{\boldsymbol
  \sigma}_{\bf k}(t)- {\boldsymbol
  \sigma}_{\bf k}^0 $ with ${\boldsymbol
  \sigma}_{\bf k}^0$ the equilibrium texture shown in Fig.~4 {\it A} of the
main article. The linearized equation reads, 
$$
\frac{\partial \delta{\boldsymbol \sigma}_{\bf k}} {\partial t}=-2 {\bf
  b}_{\bf k}^0\times   \delta{\boldsymbol \sigma}_{\bf k}-2 \delta {\bf
  b}_{\bf k}\times   {\boldsymbol \sigma}_{\bf k}^0.
$$
We define the versor $ \hat{ \bf e}_{\perp}=  \hat {\bf y}\times{\boldsymbol \sigma}_{\bf k}^0$ in the direction perpendicular to ${\bf b}_{\bf k}^0$ and the $y$ direction and pseudospin projections $\delta{\boldsymbol \sigma}_{\bf k}=\delta{\sigma}_{\bf k}^\perp \hat{ \bf e}_{\perp}+\delta{\sigma}_{\bf k}^y\hat {\bf y}$.
Axes are defined on Fig.~4 {\it B} of the main article and on
Fig.~\ref{fig:b1gtext}. For simplicity, we set the temperature to zero
and neglect collective effects and Coulomb interactions which can be
easily incorporated in the random phase approximation~\cite{and58,dev07}. 

Solving the pseudospin equation of motion for a time dependent
impulsive potential applied at $t=0$, $ v^X_{\bf k}(t)= \delta(t)\tau  \epsilon f_{\bf k}^X$ one obtains,
\begin{equation*}
  \delta \sigma_{\bf k}^\perp=-2 \sin(2 E_{\bf k} t)
  \frac{\Delta_{\bf k}}{E_{\bf k}} \tau \epsilon f_{\bf k}^X
  \end{equation*}

  \begin{equation*}
    \delta \sigma_{\bf k}^y=2 \cos(2 E_{\bf k} t)  \frac{\Delta_{\bf
      k}}{E_{\bf k}}\tau \epsilon  f_{\bf k}^X
\end{equation*}
which describe the pseudospins precessing around ${\bf b}_{\bf k}^0$.

Figures~\ref{fig:b1gtext} and 4 \textit{B} of the main article show the evolution of the texture in $B_{1g}$ and $B_{2g}$ symmetry respectively. Notice that only quasiparticles participating in the pairing have a significant time dependence and these are further restricted by the symmetry of the impulsive potential.

The coherent charge fluctuation is determined by the component of the oscillations in the $z$ direction:
\begin{equation*}
  \label{eq:nx}
\langle N_X\rangle(t)=2\sum_{\bf k}\sin(2 E_{\bf k} t) \left(\frac{f_{\bf k}^X \Delta_{\bf k}}{E_{\bf k}}\right)^2
\tau \epsilon \ \ \ \ \ \ \ t>0.
\tag{S4}
\end{equation*}
The sum runs over all the Brillouin zone. The squared factor above selects only the paired quasiparticles further restricted by the symmetry function. This shows again the selectivity of Coherent Charge Fluctuations Spectroscopy (CCFS).

The oscillation frequency depends on the energy $E_k=\sqrt{(\epsilon_k-\mu)^2+\left|\Delta_k\right|^2}$, with $\epsilon_k$ the quasiparticle energy, $\mu$ the chemical potential and $\left|\Delta_k\right|$ the superconducting gap amplitude. 
Eq.~(\ref{eq:nx}) has been evaluated numerically using a $d$-wave gap function $ \Delta_{\bf k}=\Delta_0 [\cos(k_x a)-\cos(k_y a)]/2$ with $\Delta_0=20$meV and the one band parameterization of the electronic structure of LSCO given in Ref.~\cite{Yoshida}. 
The result for $B_{2g}$ symmetry is shown in Fig.~4 \textit{C} of the main article, together with the experimental transient reflectivity oscillations visible in $B_{2g}$ geometry. We obtained a very good agreement between calculations and experiments; in particular, the oscillation frequency which corresponds to the gap amplitude close to the nodal direction is identical.

\subsection*{Effect of particle-hole excitations on the optical
  properties}
\ \

Here, we estimate the changes produced in the optical properties of the sample due to the creation of dipole allowed particle-hole (\emph{p-h}) excitations by the pump pulse. Within a mean-field picture, the optical conductivity or the charge susceptibility $\boldsymbol\chi$ of the sample changes due to: i) change of the initial state and final state occupation and ii) modification of the electronic structure due to the out-of-equilibrium distribution. We present a computation of the second effect. We assume that the main outcome of the \emph{p-h} excitations is to change the balance between the Cu ($n_d$) and O ($n_p$) hole occupation numbers putting out-of-equilibrium the charge transfer (CT) hole density $n_{CT}\equiv 2n_p-n_d$.
To take into account Cu $d_{x^2-y^2}$  and O $p_{x,y}$ orbitals, the system is described by a three-band Hubbard model:
\begin{eqnarray*}
H&=&\sum_{i}\epsilon_i \hat n_{i} +\sum_{<ij>\sigma
}t_{ij}(C_{i\sigma}^\dagger C_{j\sigma}+ h.c.)\\
&+& \sum_i U_{i} \hat n_{i\uparrow}\hat n_{i\downarrow}
+ \sum_{<ij>} U_{ij}\hat n_{i} \hat n_{j}.
\end{eqnarray*}
Here $C_{i\sigma}^\dagger$, $C_{i\sigma}$ are creation and destruction operators for holes on lattice site $i$ with spin $\sigma$ and we defined the occupation number operators, $\hat n_{i\sigma}=C_{i\sigma}^\dagger C_{i\sigma}$, $\hat n_{i}= \hat n_{i\uparrow}+\hat n_{i\downarrow}$. The on-site energies are $\epsilon_i=\epsilon_p$ ($\epsilon_d$) and the on-site
repulsions, $U_i=U_p$ ($U_d$) for $i$ in an O (Cu) site; and the hopping matrix elements, $t_{ij}=\pm t_{pd} (\pm t_{pp}$) for Cu-O (O-O) nearest neighbor sites, with the sign depending on the relative phase of the Wannier orbitals involved. For the inter-site interaction, we keep only the nearest neighbor repulsion $U_{ij}=U_{pd}$ between Cu and O. We define the CT gap $\Delta\equiv \epsilon_p-\epsilon_d$. We take the same parameters as in Ref.~\cite{Lorenzana2003}, 
 except for $\Delta=3eV$. This ensures that the first maximum in the charge transfer band in the optical conductivity of the parent compound is at 2.16 eV, in good agreement with the experiments~\cite{Uchida1991}, instead of a higher value obtained with the first principle
 parameters of Ref.~\cite{Lorenzana2003}. 

To account for the strong correlations on Cu, we treat the $d$ orbital using the Gutzwiller approximation and the other interactions using the Hartree-Fock approximation. This leads to a single particle Hamiltonian with renormalized energies, which at equilibrium depend self-consistently on the charge distribution. Our strategy is to compute the optical conductivity at mean-field level in the ground state and in the presence of the out-of-equilibrium charge distribution. We also neglect any asymmetry in this out-of-equilibrium distribution. In other words we assume that the out-of-equilibrium distribution has $A_{1g}$ symmetry.

The leading effect of the out-of-equilibrium charge distribution is to renormalize the effective energies at mean-field level. Since this happen through total charges that are integrals of the distribution, this effect will be rather insensitive to its precise form. The charge
is put out-of-equilibrium by adding a constraint implemented through a Lagrange multiplier. Thereafter, the optical conductivity and the charge susceptibility $\boldsymbol\chi$ are obtained at mean-field level, and the derivative with respect to the out-of-equilibrium charge is
evaluated numerically. For the doped case, the ground-state is assumed to consist of stripes, as in Ref.~\cite{Lorenzana2003}, which gives a good overall agreement with experiments for the optical conductivity.

Figure~\ref{fig:deps2dnct} shows the result of the computation for the transient imaginary part of the in-plane dielectric function obtained as $$\delta \epsilon_{xx}(\omega,t)=- 4\pi
\frac{\partial{\chi_{xx}}}{\partial n_{CT}}(\omega) \delta n_{CT}(t).$$ with $\delta n_{CT}(t)= n_{CT}(t)- n_{CT}^0$ and $n_{CT}^0 $ being the equilibrium charge transfer density.

Given the simplifications in the computation, we obtain a fair overall
agreement with the experiment adjusting only the parameter $\delta
n_{CT}(500 fs)=8\cdot10^{-3}$ to fit the intensity.  This gives the
estimate of the number of transferred holes per Cu mentioned in the
main article. 

The number of excited electrons per unit cell may be estimated independently in terms of experimental parameters using the following formula:
\begin{equation*}
	n=2 A V\frac{F}{l_s^2 \Delta E}\int^{l_s}_{0}e^{-z/l_s}dz
\end{equation*}
where $A$ is the absorption coefficient and $l_s$ the penetration depth, both at the pumping wavelength, $V$ the unit cell volume, $F$ the pumping fluence in $J/m^2$ and $\Delta E$ the transition energy. For a fluence of 300 $\mu J/cm^2$, we obtain a value of $n=4\cdot10^{-3}$ per Cu atom, in fair agreement with the theoretical calculation of excited electrons $\delta n_{CT}$.

$\delta n_{CT}(t)$ is positive, which is consistent with the fact that in the ground state there are far more holes on Cu than on O atoms, so the laser pulse will tend to decrease $n_d$ and increase $n_p$. The transient dielectric function has a strong energy dependence due to transfer of spectral weight among the different absorption bands which leads to regions of positive and negative transient reflectivity as shown in Fig.~2 of the main article.

The profile of $\epsilon_2$ can be understood as due to a softening of the charge transfer edge due to a downward renormalization of the mean-field charge transfer energy $\tilde\Delta$. The leading effect is due to the nearest neighbor repulsion $\delta\tilde\Delta\sim -2 U_{pd}\delta n_{CT}$. The structures appear at higher energy than in the experiment probably because our mean-field approach underestimates the softening of the CT edge with doping.

The same strategy can be used to compute the electronic Raman tensors defined in the main text and in the previous section.

\section{Legends for Supporting Movies}
\indent

{\bf  Supporting Movie 1. La A$_{1g}$ phonon.} The movie shows the atomic motions during the La A$_{1g}$ phonon; La and O atoms are moving along the $c$-axis of La$_{2-x}$Sr$_x$CuO$_4$ unit cell. The amplitude of atomic motions is exaggerated for clarity.
~\\

{\bf  Supporting Movie 2. Charge fluctuation in $B_{2g}$ geometry}. The movie shows the theoretical computation of the charge fluctuation in $B_{2g}$ geometry and the time evolution of the pseudospins
encoding BCS wave function. Notations are the same as in Figs. 4 \textit{B} and~\ref{fig:b1gtext}. For $t<0$ pseudospins form the equilibrium texture characteristic of a $d$-wave superconductor. At $t=0$ the Raman impulsive field of the pump pulse is applied (red arrows). For $t>0$ pseudospins precess around their equilibrium position. The vertical projections of the pseudospins determine the charge fluctuation which decays as pseudospins in different regions of energy and momentum lose coherence in time.


\begin{thebibliography}{32}
%


\bibitem{BCS}
Bardeen J, Cooper LN, Schrieffer JR (1957) {\em Theory of Superconductivity.} \textit{Phys. Rev.} 108:1175-1204.

\bibitem{max50} Maxwell E (1950) \em{Isotope effect in the superconductivity of mercury.} \textit{Phys. Rev.} 78:477.

\bibitem{rey50} Reynolds CA, Serin B, Wright WH, Nesbitt LB (1950) \em{Superconductivity of isotopes of mercury.} \textit{Phys. Rev.} 78:487.

   \bibitem{ber1986}
Bednorz JG, M\"uller KA (1986) {\em Possible high-$T_c$ superconductivity in the Ba-La-Cu-O system.} \textit{Zeitschrift f\"ur Physik B Condensed Matter} 64:189-193.


\bibitem{ore12}
Orenstein J (2012) {\em Ultrafast spectroscopy of quantum materials.} \textit{Phy. Today} 65:44.

   \bibitem{Perfetti2007}
Perfetti L, et al. (2007) {\em Ultrafast Electron Relaxation in Superconducting Bi$_2$Sr$_2$CaCu$_2$O$_{8+\delta}$ by Time-Resolved Photoelectron Spectroscopy.} \textit{Phys. Rev. Lett.} 99:197001.


\bibitem{how04} Howell PC, Rosch A, Hirschfeld PJ (2004) {\it Relaxation of hot quasiparticles in a $d$-Wave superconductor.} \textit{Phys. Rev. Lett.} 92:037003.







   \bibitem{Kaindl2000}
Kaindl RA, et al. (2000) {\em Ultrafast Mid-Infrared Response of YBa$_2$Cu$_3$O$_{7- \delta}$.} \textit{Science} 287:470-473.

\bibitem{sto11} Stojchevska L, et al. (2011) {\it Mechanisms of nonthermal destruction of the superconducting state and melting of the charge-density-wave state by femtosecond laser pulses.} \textit{Phys. Rev. B} 84:180507.

\bibitem{Pashkin2010} Pashkin A, et al. (2010) \em{Femtosecond response of quasiparticles and phonons in superconducting  YBa$_{2}$Cu$_3$O$_{7- \delta}$ studied by wideband terahertz spectroscopy.} \textit{Phys. Rev. Lett.} 105:067001.

\bibitem{Beck2011} Beck M, et al. (2011) \em{Energy-Gap Dynamics of Superconducting NbN Thin Films Studied by Time-Resolved Terahertz Spectroscopy.} \textit{Phys. Rev. Lett.} 107:177007.

\bibitem{ged04}
Gedik N, et al. (2004) {\em Single-quasiparticle stability and quasiparticle-pair decay in YBa$_2$Cu$_3$O$_{6.5}$.} \textit{Phys. Rev. B} 70:014504.

\bibitem{nuh} Gedik N, et al. (2005) \em{Abrupt Transition in Quasiparticle Dynamics at Optimal Doping in a Cuprate Superconductor System.} \textit{Phys. Rev. Lett.} 95:117005.

\bibitem{kab05}
Kabanov VV, Demsar J, Mihailovic D (2005) {\em Kinetics of a superconductor excited with a femtosecond optical pulse.} \textit{Phys. Rev. Lett.} 95:147002.

\bibitem{Giannetti2011} 
 Giannetti C, et al. (2011) \em{Revealing the high-energy electronic excitations underlying the onset of high-temperature superconductivity in cuprates.} \textit{Nature Comm.} 2:353.
 
\bibitem{cor11} Cort\'es R, et al. (2011) \em{Momentum-Resolved Ultrafast Electron Dynamics in Superconducting Bi$_2$Sr$_2$CaCu$_2$O$_{8+\delta}$.} \textit{Phys. Rev. Lett.} 107:097002.


\bibitem{Carbone2008}
Carbone F, Yang DS, Giannini E, Zewail AH (2008) {\em Direct role of structural dynamics in electron-lattice coupling of superconducting cuprates.} {\it Proc Natl Acad Sci USA} 105:20161-20166.

\bibitem{carbonereview} 
Carbone F, Gedik N, Lorenzana J, Zewail AH, (2010) {\em Real-Time Observation of Cuprates Structural Dynamics by Ultrafast Electron Crystallography.} \textit{Adv. Cond. Matt. Phys.} 2010:958618.

\bibitem{Merlin} Merlin R (1997) \em{Generating coherent $THz$ phonons with light pulses.} \textit{Solid State Comm.} 102:207-220.

\bibitem{Merlin2002} Stevens TE, Kuhl J, Merlin R (2002) \em{Coherent phonon generation and the two stimulated Raman tensors.} \textit{Phys. Rev. B} 65:144304.

\bibitem{Zeiger} Zeiger HJ, et al. (1992) \em{Theory for displacive excitation of coherent phonons.} \textit{Phys. Rev. B} 45:768.

\bibitem{Mazin1994} Mazin II, Liechtenstein AI, Jepsen O, Andersen OK, Rodriguez CO (1994) \em{Displacive excitation of coherent phonons in YBa$_2$Cu$_3$O$_{7}$.} \textit{Phys. Rev. B} 49:9210.

\bibitem{rif07}
Riffe DM, Sabbah AJ (2007) {\em Coherent excitation of the optic
  phonon in Si: Transiently stimulated Raman scattering with a
  finite-lifetime electronic excitation.}  \textit{Phys. Rev. B} 76:085207.

\bibitem{Albrecht1992} Albrecht W, Kruse Th, Kurz H (1992), \em{Time-resolved observation of coherent phonons in superconducting YBa$_2$Cu$_3$O$_{7-\delta}$ thin films.} \textit{Phys. Rev. Lett.} 69:1451.

\bibitem{mis02}
Misochko OV, Georgiev N, Dekorsy T, Helm M (2002) {\em
 Two crossovers in the pseudogap regime of YBa$_2$Cu$_3$O$_{7-\delta}$ superconductors observed by ultrafast spectroscopy.} \textit{Phys. Rev. Lett.} 89:067002. 
 
 \bibitem{zha04}
Zhao J, Bragas AV, Lockwood DJ, Merlin R (2004) {\em Magnon squeezing in an antiferromagnet: Reducing the spin noise below the standard quantum limit.} \textit{Phys. Rev. Lett.} 93:107203.

\bibitem{bao04}
Bao JM, Pfeiffer LN, West KW, Merlin R (2004) {\em Ultrafast dynamic control of spin and charge density oscillations in a GaAs quantum well.} \textit{Phys. Rev. Lett.} 92:236601.

\bibitem{and58} Anderson PW (1958) \em{Random-Phase Approximation in the theory of superconductivity.} \textit{Phys. Rev.} 112:1900-1916.

\bibitem{war05} Warner GL, Leggett AJ (2005) \em{Quench dynamics of a superfluid Fermi gas.} \textit{Phys. Rev. B} 71:134514.

\bibitem{pseudospins_ref} The  pseudospin operators are defined as
${\sigma}^x_{\bf k}=(c_{{\bf k}\uparrow}c_{-{\bf k}\downarrow} +h.c.)$,
$i{\sigma}^y_{\bf k}=(c_{{\bf k}\uparrow}c_{-{\bf k}\downarrow} -h.c.)$,
${\sigma}^z_{\bf k}=1-n_{{\bf k}\uparrow}-n_{-{\bf
    k}\downarrow}$. Here $n_{{\bf
    k}\sigma}=c^\dagger_{{\bf k}\sigma} c_{{\bf k}\sigma}$ and
 $c^\dagger_{{\bf k}\sigma}$ ($c_{{\bf k}\sigma}$) are creation
 (destruction) operators for electrons.

    
\bibitem{sch96} Slichter CP (1996) \em{Principles of Magnetic Resonance (Springer Series in Solid-State Sciences) (v. 1)}.

\bibitem{eli60} Eliashberg GM (1960) {\em Interactions between electrons and lattice vibrations in a superconductor.} \textit{Soviet Physics JETP} 11:696.

\bibitem{and07} Anderson PW (2007) \em{Is there a glue in cuprate superconductors?} \textit{Science} 316:1705-1707.

\bibitem{zaa11} Zaanen J (2011) {\em  A modern, but way too short history of the theory of superconductivity at a high temperature}, in {\em 100 Years of Superconductivity.} Rogalla H and Kes PH (eds.), (Taylor and Francis).

\bibitem{and87} Anderson PW (1987) {\em The Resonating Valence Bond State in La$_2$CuO$_4$ and Superconductivity.} \textit{Science} 235:1196-1198.

\bibitem{lu} Mansart B, et al. (2012) \em{Evidence for a Peierls phase-transition in a three-dimensional multiple charge-density waves solid.} \textit{Proc. Natl. Acad. Sci. U.S.A.} 109:5603-5608.

\bibitem{Sugai2003} Sugai S, Suzuki H, Takayanagi Y, Hosokawa T, Hayamizu N (2003) \em{Carrier-density-dependent momentum shift of the coherent peak and the LO phonon mode in $p$-type high-T$_c$ superconductors.} \textit{Phys. Rev. B} 68:184504.

\bibitem{dev07} Devereaux TP, Hackl R (2007) \em{Inelastic light scattering from correlated electrons.} \textit{Rev. Mod. Phys.} 78:175.

\bibitem{Uchida1991} Uchida S, et al. (1991) \em{Optical spectra of La$_{2-x}$Sr$_x$CuO$_4$: Effect of carrier doping on the electronic structure of the CuO$_2$ plane.} \textit{Phys. Rev. B} 43:7942.

\bibitem{Lorenzana2003} Lorenzana J, Seibold G (2003) \em{Dynamics of Metallic Stripes in Cuprates.} \textit{Phys. Rev. Lett.} 90:066404.

\bibitem{raman}
 Charge fluctuations are defined by the operator
$
N_X=\sum_{{\bf k}\sigma} f_{\bf k}^X n_{{\bf k}\sigma},
$
with $f_{\bf k}^{B_{1g}}=[\cos(k_x a)-\cos(k_y a)]/2$, $f_{\bf
  k}^{B_{2g}}=\sin(k_x a)\sin(k_y a)$. The Raman time dependent
potentials produced by the pump electric field, ${\bf E}$, are
$v^X_{\bf k}(t)=f^X_{\bf  k}v_X(t)$ with,
$
v_X(t)=-\frac12{\bf E}(t).\partial {\boldsymbol
    \chi}(\omega_{L}) /{\partial \langle  N_X\rangle}.{\bf E}(t).
$ with ${\boldsymbol \chi}$ the charge susceptibility and its density
derivative the conventional Raman tensor (see SI Text).

\bibitem{mai08} Maier TA, Poilblanc D, Scalapino DJ (2008) \em{Dynamics of the pairing interaction in the Hubbard and t-J models of high-temperature superconductors.} \textit{Phys. Rev. Lett.} 100:237001.

\bibitem{sca95} Scalapino DJ (1995) {\em The case for $d_{x^2-y^2}$ pairing in the cuprate superconductors.} \textit{Physics Reports} 250:329-365.
    
\bibitem{pin97} Pines D (1997) \em{Nearly antiferromagnetic Fermi liquids: a progress report.} \textit{Zeitschrift f\"ur Physik B} 103:129-135.   


\bibitem{Cardona} Cardona M (1982) \em{Light Scattering in Solids II: Basic Concepts
  and Instrumentation}, Springer-Verlag.
  
    \bibitem{Parmenter} For the definitions of the pseudospins we follow
    the conventions of Parmenter RH (1965) \em{Time- and
      position-dependent superconductivity.} \textit{Phys. Rev.}
    137:A161-A163. 
    
      \bibitem{Yoshida} Yoshida T, et al. (2006) \em{Systematic doping evolution of the underlying Fermi surface of La$_{2-x}$Sr$_x$CuO$_4$.} \textit{Phys. Rev. B} 74:224510.
      
\end{thebibliography}
\end{document}